\documentclass[preprint]{aastex}

\usepackage{savesym}
\usepackage{SIunits}
\restoresymbol{SI}{square}
\usepackage{graphicx}
\usepackage{natbib}

\addunit{\gauss}{G}
\addunit{\parsec}{pc}
\addunit{\yr}{yr}
\addunit{\erg}{erg}
\addunit{\eV}{eV}

\newcommand{\Gyr}{\giga\yr}
\newcommand{\Myr}{\mega\yr}

\newcommand{\muG}{\micro\gauss}
\newcommand{\kpc}{\kilo\parsec}
\newcommand{\cmsps}{\centi\meter\squared\;\reciprocal\second}
\newcommand{\kmps}{\kilo\meter\;\reciprocal\second}

\shorttitle{Global Cosmic-Ray-Driven Galactic Dynamo}
\shortauthors{M. Hanasz et al.}

\received{\ldots}
\accepted{\ldots}

\begin{document}

\title{Global galactic dynamo driven by cosmic rays and exploding magnetized stars}

\author{Micha\l{} Hanasz, Dominik W\'olta\'nski, Kacper Kowalik}

\altaffiltext{1}{Centre for Astronomy, Nicholas Copernicus University,
  PL-87148 Piwnice/Toru\'n, Poland; mhanasz@astri.uni.torun.pl}

\begin{abstract}
We report first results of first global galactic-scale CR-MHD simulations of cosmic-ray-driven dynamo. We investigate the dynamics of magnetized interstellar medium (ISM), which is   dynamically coupled with the cosmic-ray (CR) gas.  We assume that exploding stars deposit  small-scale,  randomly oriented, dipolar magnetic fields into the differentially rotating ISM, together with a portion of cosmic rays, accelerated in supernova shocks.   We conduct numerical simulations with the aid of a new parallel MHD code PIERNIK.  We find that the initial magnetization of galactic disks by exploding magnetized stars forms favorable conditions for the cosmic-ray-driven dynamo.   We demonstrate that dipolar magnetic fields supplied on small  supernova remnant scales can be amplified exponentially by the CR-driven dynamo, to the present equipartition values, and transformed simultaneously to large galactic scales. The resulting magnetic field structure in an evolved galaxy appears spiral in the face-on view and reveals the so-called X-shaped structure in the edge-on view. 
\end{abstract}


\section{Introduction}

It is commonly believed that galactic magnetic fields are generated by a kind of dynamo process. Dynamos  require, however seed fields. Various cosmological processes have been considered as candidates for the seed field generation mechanisms, such as phase transitions in the early universe, and battery effects on protogalactic scales.  However, known processes of this kind lead to weak fields of the order of  $\unit{\power{10}{-20}}{\gauss}$ \cite*[see][]{2002RvMP...74..775W}.

Another possibility, suggested by \cite*{syrovatskii-70}, developed by \cite*{1973SvA....17..137B}, and by \cite*{1987QJRAS..28..197R,2006AN....327..395R}, is that the seed fields may have been created during early stages of galactic evolution by a stellar-scale Biermann  battery, subsequently amplified by stellar dynamos, and then spread into the interstellar medium (ISM) by stellar explosions. \cite*{1987QJRAS..28..197R} estimates that a contribution of $10^6$ Crab-type, randomly oriented plerionic supernova remnants may lead to $\unit{\power{10}{-9}}{\gauss}$ mean magnetic field on galactic scales. 

In a series of recent papers \citep{2004ApJ...605L..33H,2006AN....327..469H,2009A&A...498..335H}, we have shown  that in local, shearing-box MHD simulations, weak magnetic fields can be
amplified  up to a few $\muG$ within $\unit{1-2}{\Gyr}$, by the dynamo process driven by buoyancy of cosmic rays, as  proposed by \cite{1992ApJ...401..137P}. The aim of this paper is to demonstrate that small-scale, randomly oriented, dipolar magnetic fields of stellar origin, can be amplified in a similar rate, and regularized by the cosmic-ray-driven dynamo in global galactic disk CR--MHD simulations.

\section*{Global model of the CR-driven dynamo}

To built up a global galactic disk model, we use the realistic Milky Way gravitational potential by \cite*{1991RMxAA..22..255A}. We neglect, however, the galactic bulge contribution, to obtain smoother rotation curve near the galactic center. Without the bulge contribution rotational velocity approaches $V_\phi \simeq \unit{220}{\kilo\meter\per\second}$ at the galactocentric radius of Sun.
The initial setup for the gaseous disk,  supernova rate, and its spatial distribution is based on the global ISM model by \cite*{1998ApJ...497..759F}. 

The global model of CR-driven dynamo involves the following
elements, introduced previously in the local models:
(1) the cosmic ray component -- a relativistic gas described by the
diffusion--advection transport equation, supplemented to the standard set of
resistive MHD equations.
(2) Cosmic rays supplied in supernova remnants. The cosmic ray input of individual SNe is assumed $10\%$ of the typical SN kinetic energy output ($=\unit{\power{10}{51}}{\erg}$), 
while  the thermal energy output from supernovae is neglected.
(3) Following \cite*{1999ApJ...520..204G}, we assume that CRs  diffuse anisotropically along magnetic field lines.
(4) We incorporate a finite resistivity of the ISM,  
to permit topological evolution of galactic magnetic fields via anomalous resistivity processes 
\citep[see][and references therein]{2002A&A...386..347H}, 
and/or via turbulent reconnection \citep{2009ApJ...700...63K}, on small spatial scales, which are unresolved in our simulations.
(5) Differential rotation of the interstellar gas, which currently follows the assumed form of galactic gravitational potential.

Moreover, contrary to our previous models, we assume that
(6) no magnetic field is present in the initial configuration, and that a weak, randomly oriented, dipolar magnetic field is supplied in 10 \% of supernova remnants. 

\section{Numerical setup}
\begin{figure*}[ht]
  \centerline{\includegraphics[width=0.35\textwidth]{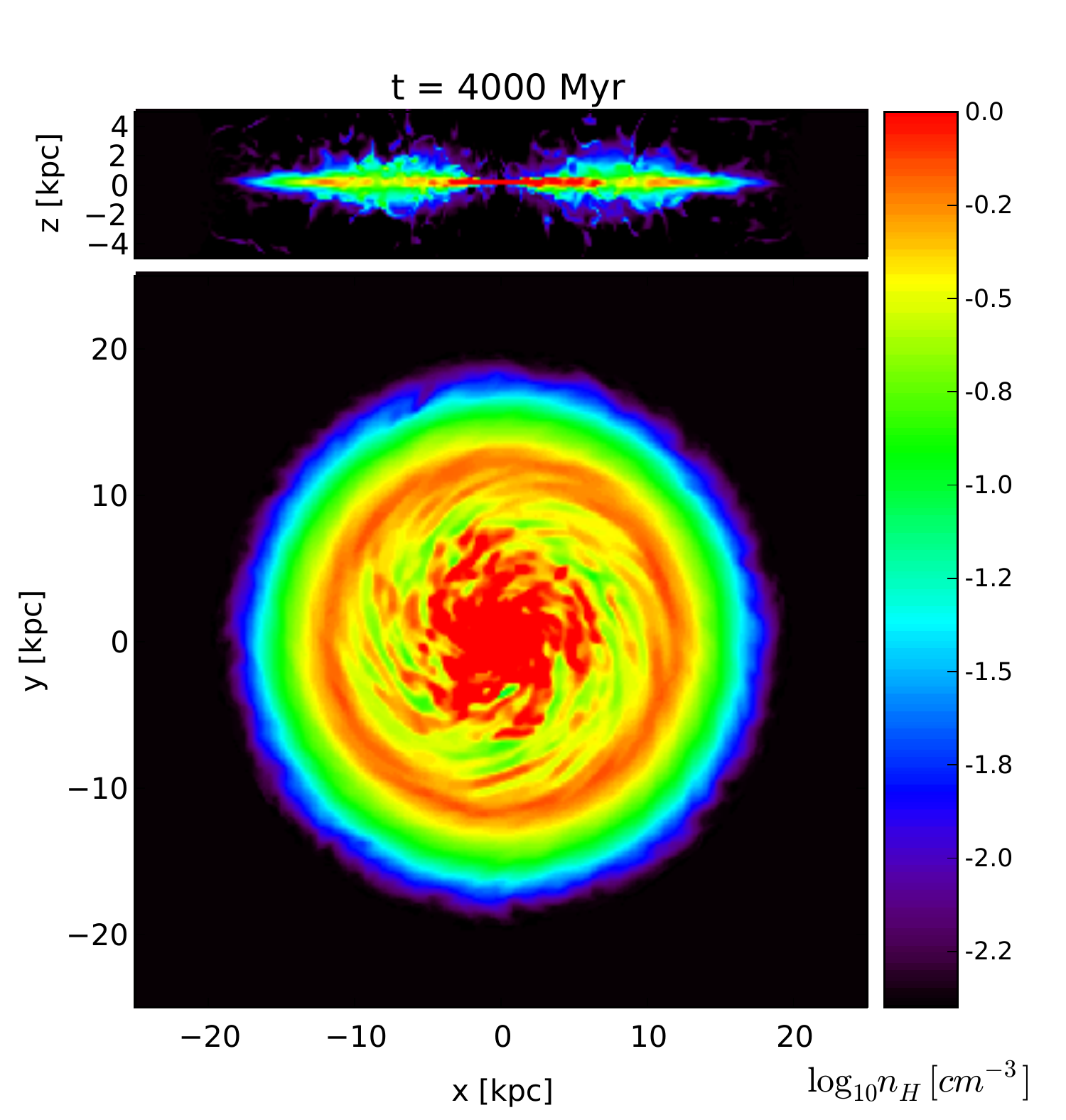}
              \includegraphics[width=0.35\textwidth]{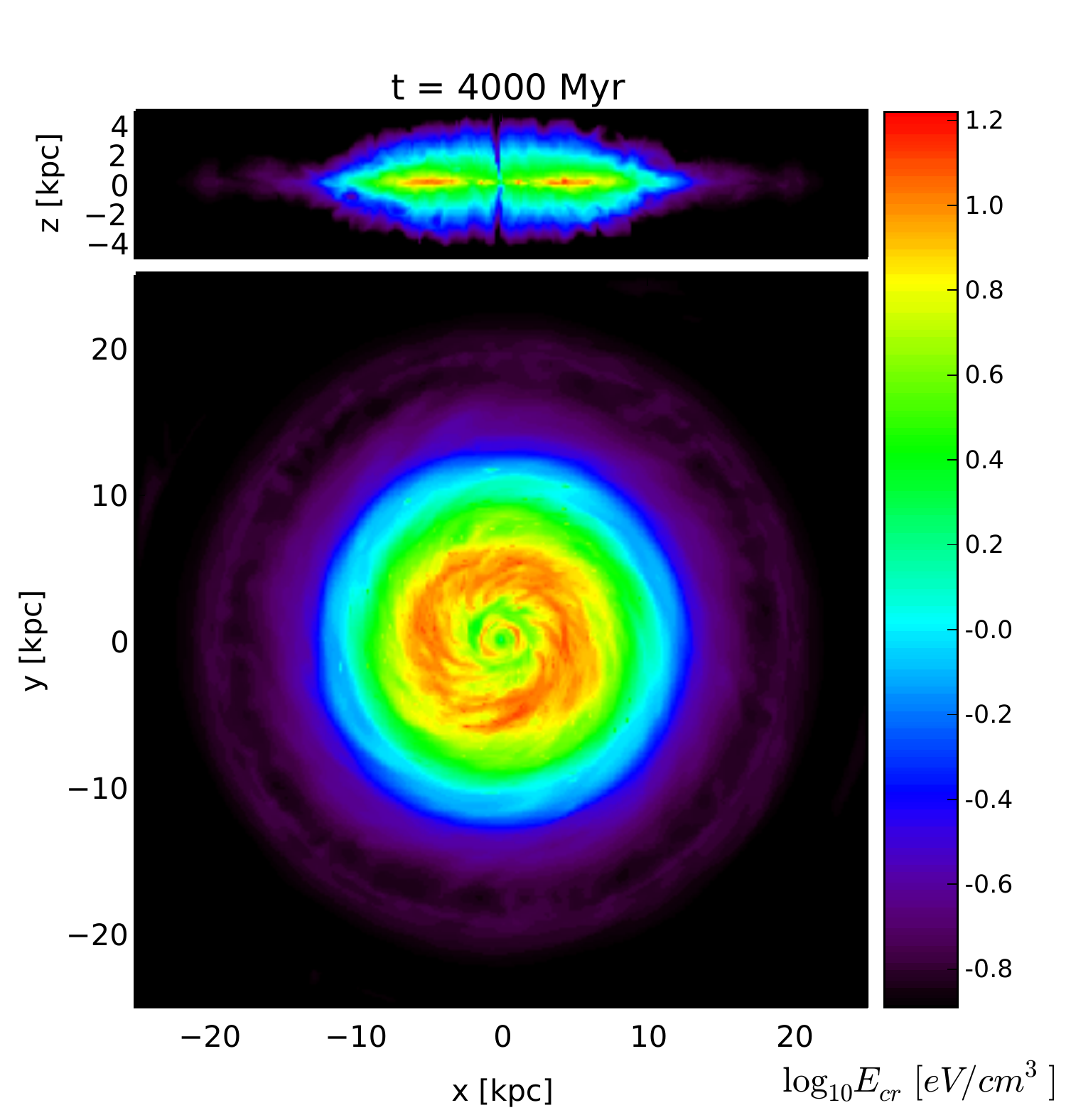}}
\caption{Logarithm of gas number density in units of hydrogen atom per cubic centimeter (left) and cosmic ray energy density distribution in units of electron volt per cubic centimeter (right) at $t= \unit{4}{\Gyr}$.}
\label{fig:slices-de}
\end{figure*}
%
%
%
\begin{figure*}[ht]
  \centerline{\includegraphics[width=0.35\textwidth]{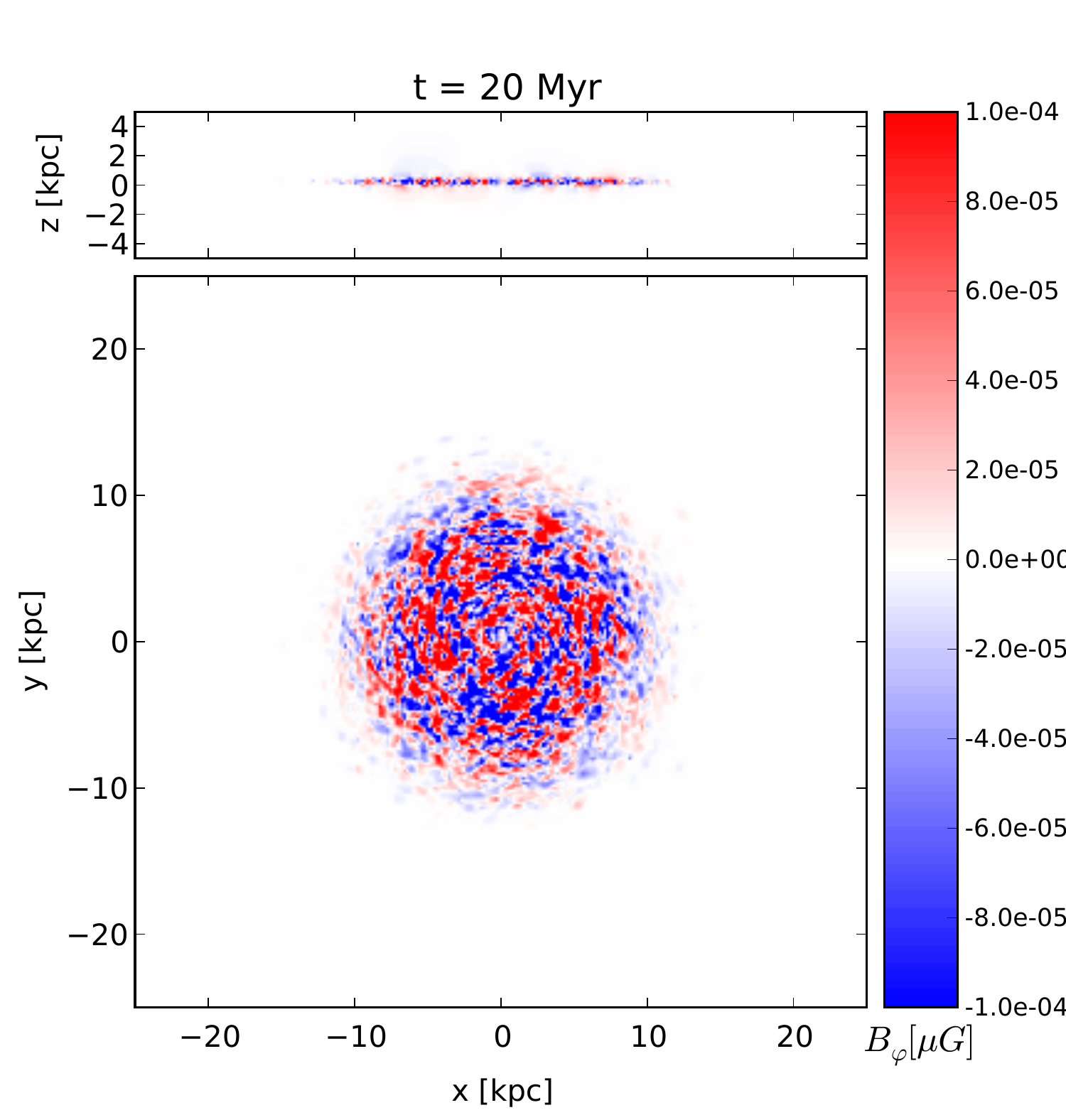}
              \includegraphics[width=0.35\textwidth]{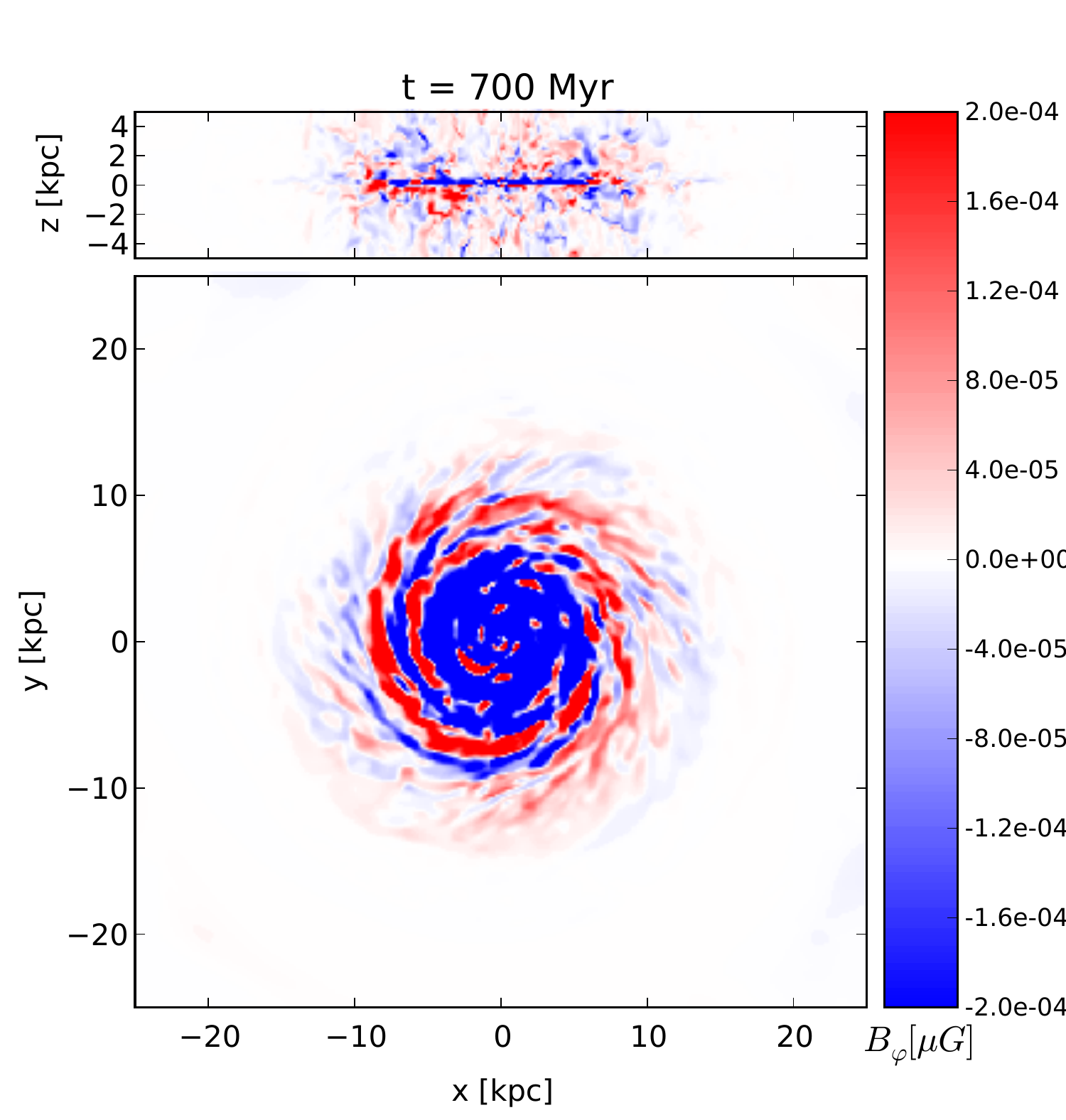}}
  \centerline{\includegraphics[width=0.35\textwidth]{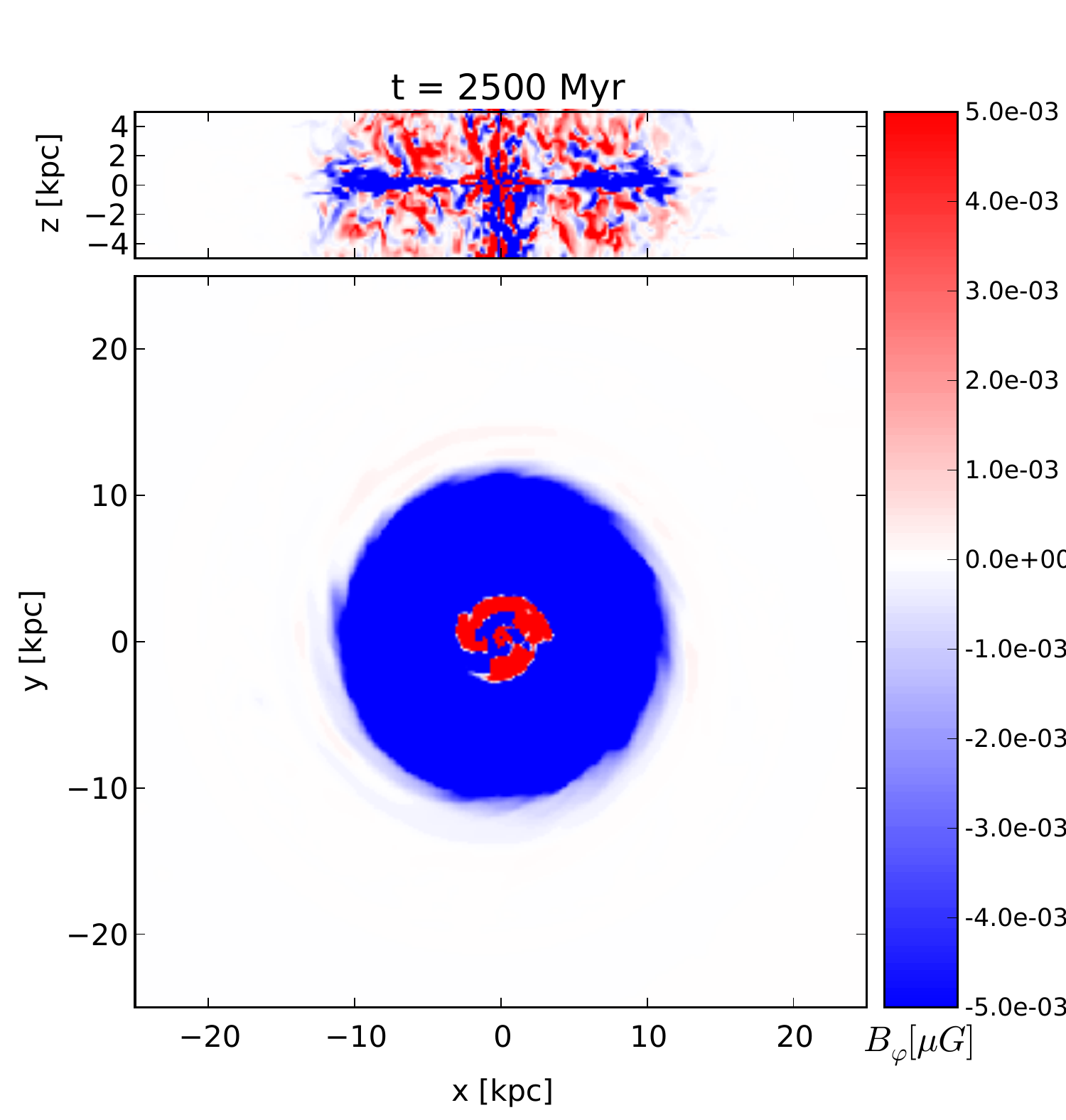}
              \includegraphics[width=0.35\textwidth]{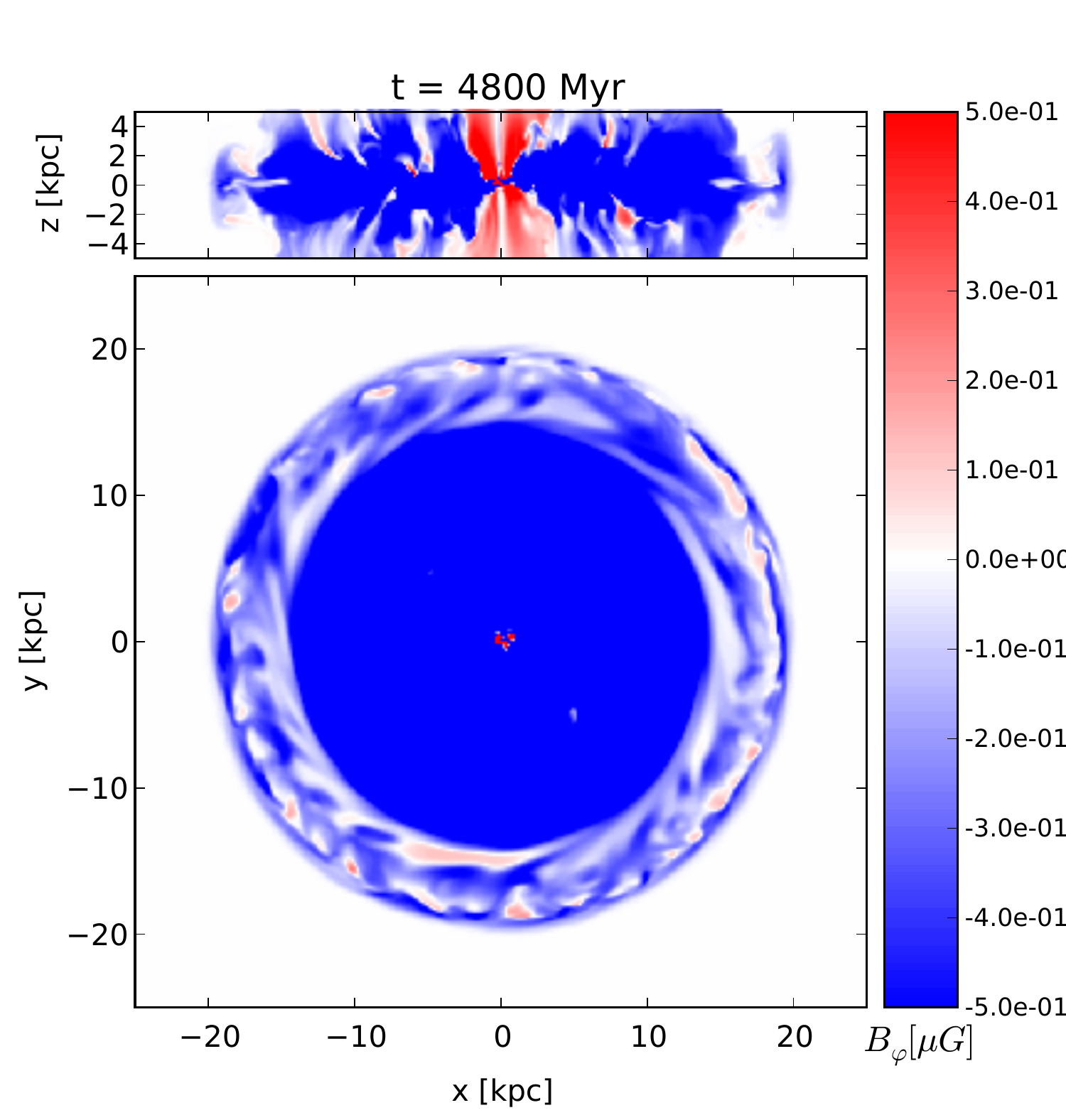}} 
\caption{Distribution of toroidal magnetic field at $t=\unit{20}{\Myr}$ (top left), $t=\unit{700}{\Myr}$ (top right), $t=\unit{2.5}{\Gyr}$ (bottom left), 
and $t = \unit{4.8}{\Gyr}$ (bottom right). Unmagnetized regions of the volume are white, while positive and negative toroidal magnetic fields are marked with red and blue, respectively. Note that the color scale in magnetic field maps is saturated to enhance weaker magnetic field structures in disk peripheries.
The maximum magnetic field strengths are $5.9 \cdot 10^{-4}$, $4.4 \cdot 10^{-3}$, $1.5$, and $29$ $\muG$ at $t=0.02$, $0.7$,  $2.5$, and $4.8 \Gyr$, respectively.}
\label{fig:slices-bb}
\end{figure*}				
%
%
%
\begin{figure}[ht]
\centerline{\includegraphics[width=0.5\columnwidth]{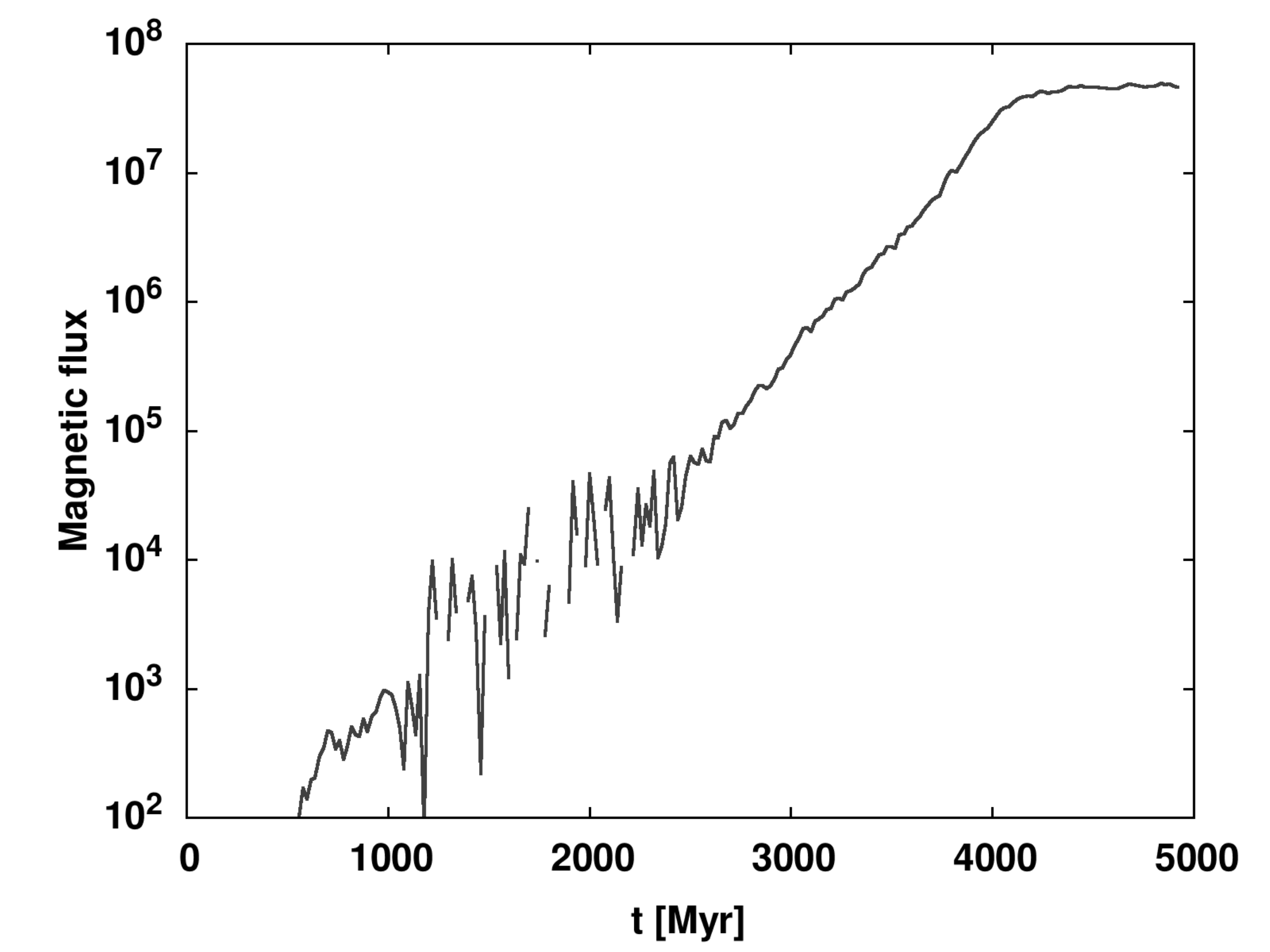} \quad 
	    \includegraphics[width=0.5\columnwidth]{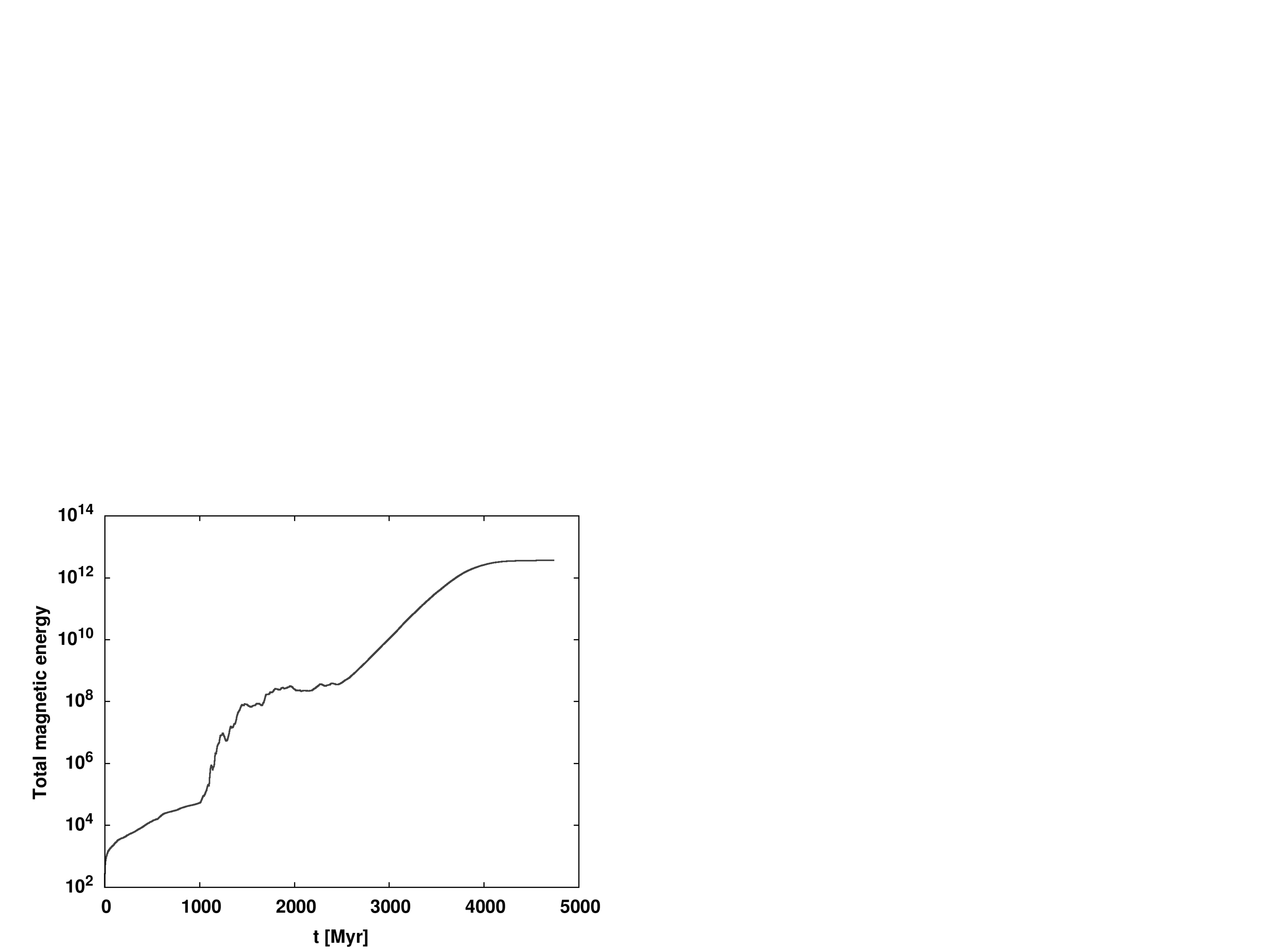}} 
\caption{Temporal evolution of toroidal magnetic flux  and total magnetic energy in scaled units. The final saturation level corresponds to the equipartition magnetic fields.}
\label{fig:mflx-emag}
\end{figure}				
%
%
%
\begin{figure}[ht]
\centerline{\includegraphics[width=0.6\columnwidth]{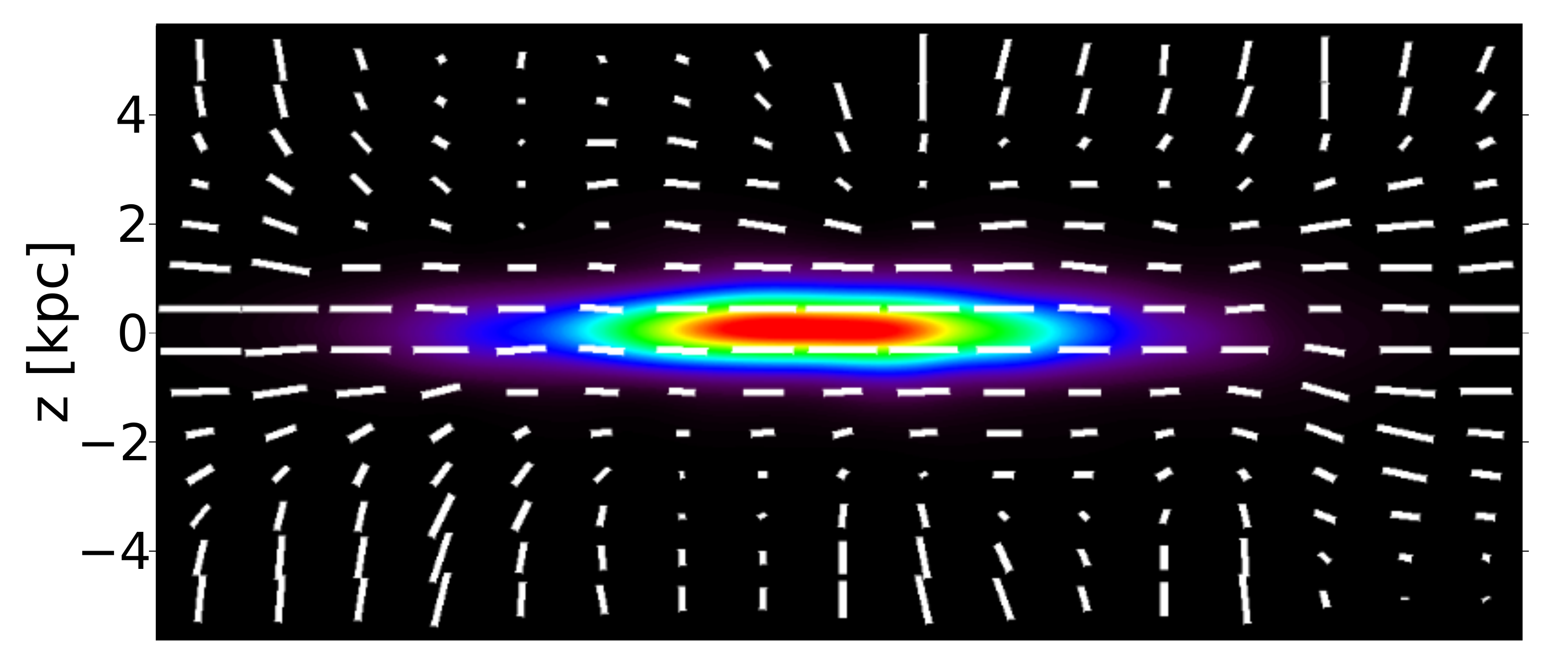}}
\centerline{\includegraphics[width=0.6\columnwidth]{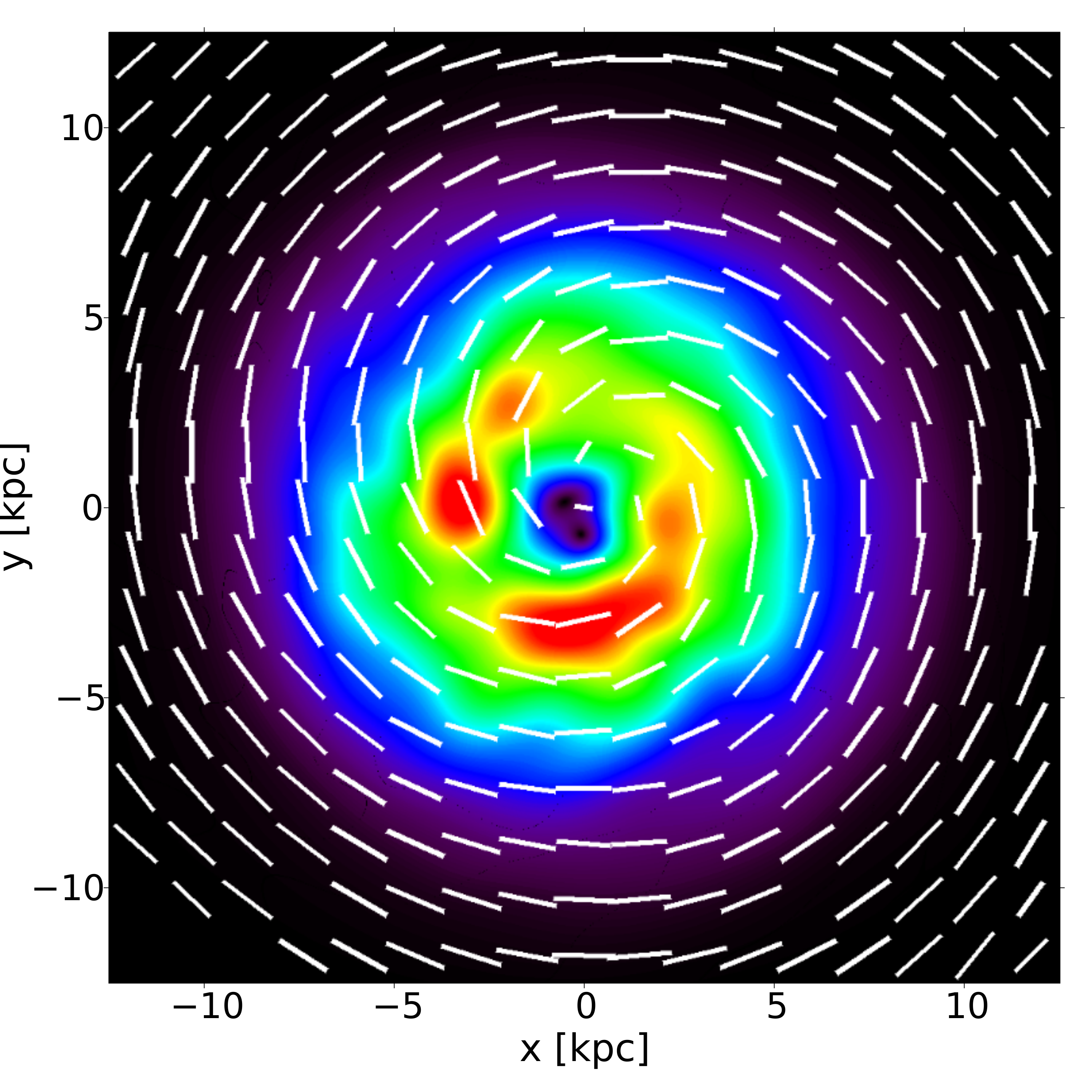}}
\caption{
Synthetic radio maps of polarized intensity (PI) of synchrotron emission, together with polarization vectors are shown for the edge-on and face-on views of the galaxy at $t=\unit{4.8}{\Myr}$. 
Vectors direction resembles electric vectors rotated by $90^{\circ}$, and their lengths are proportional to the degree of polarization.
}
\label{fig:radiomaps}
\end{figure}

We use the  PIERNIK MHD code \citep{2008arXiv0812.2161H,2008arXiv0812.2799H}, which  is a grid-MHD code based on the Relaxing TVD (RTVD) scheme by \cite*{jin-xin-95} and \cite*{2003ApJS..149..447P}. PIERNIK is parallelized by means of block decomposition with the aid of MPI library.  The original scheme is extended to deal with dynamically independent, but interacting fluids: thermal gas and a diffusive cosmic ray gas, described within the fluid  approximation \citep{2008arXiv0812.4839H}. The code has been equipped with the resistivity module \citep{2009arXiv0901.0104H}. The  induction equation, including the Ohmic resistivity term, is integrated with the aid of the constraint transport (CT) algorithm \citep{1988ApJ...332..659E}. 

The CR diffusion algorithm follows the implementation of CR transport described by \cite*{2003A&A...412..331H}.  The  CR diffusion--advection equation is integrated in a conservative manner. The CR anisotropic diffusion is parameterized by parallel and perpendicular (with respect to the local magnetic field direction) diffusion coefficients $K_\parallel$ and $K_\perp$.  We assume that CRs couple to gas through the CR pressure gradient term in the gas equation of motion \citep[see][]{1990acr..book.....B}.

The simulation presented here has been performed with the resolution of $500 \times 500 \times 100$ grid cells, distributed in $10 \times 10 \times 2$ equal size MPI blocks, in the Cartesian domain spanning the volume $\unit{50}{\kpc}\times\unit{50}{\kpc}\times\unit{10}{\kpc}$ in $x$, $y$ and $z$ directions respectively. Initial gas distribution represents a hydrostatic equilibrium state for gas column density given by \cite{1998ApJ...497..759F}. The probability distribution for occurrence of SN remnants, derived from  \cite*{1998ApJ...497..759F} was fixed over simulation time. Magnetic vector potential of a single SNR  is computed, following \cite{Jackson-99} for a current-loop of a radius comparable to the radius of SNR (\unit{50}{\parsec} in the present model). We stop the dipolar magnetic field input after \unit{1}{\Gyr} of the disk evolution, because the corresponding procedure is computationally expensive, while the contribution of the supplied dipoles becomes negligible with respect to dynamo amplification  dominating at later times. The values of diffusion coefficients adopted for the present
simulation are $K_\parallel = \unit{3\times\power{10}{28}}{\cmsps}$ and 
$K_\perp= \unit{3 \times \power{10}{26}}{\cmsps}$ \citep[see][and references therein for a more extensive discussion on CR diffusion]{2007ARNPS..57..285S}. In the present model we assume a uniform magnetic diffusivity  $\eta=\unit{3\cdot\power{10}{24}}{\cmsps}$, and neglect heating of the thermal gas by resistive dissipation of magnetic fields.  We apply outflow boundary conditions on upper and lower domain boundaries, and enforce gas to follow the prescribed rotation curve outside the outer disk radius $r_{\rm max} = \unit{21.5}{\kpc}$. To prevent significant mass losses, due to the CR-driven wind, we re-supply into the disk the gas lost from the computational domain. 
\section{Results}
\subsection{Structure of ISM}
The gas density distribution and cosmic ray energy distribution in the disk are displayed in vertical and horizontal slices through the disk center, for $t=\unit{4}{\Gyr}$, in Figure~\ref{fig:slices-de}, while the distribution of toroidal magnetic field component is shown for four different epochs in Figure~\ref{fig:slices-bb}.  

The gas and cosmic ray distributions achieve dynamical equilibrium in about $\unit{100}{\Myr}$ of initial SN activity. The characteristic shape of the disk, apparent in the gas and CR distributions, reflects the initial ring-like structure of the gaseous disk. Vertical wind blowing with the speed exceeding $\unit{100}{\kmps}$, apparently above and below the central part of the disk, carries out the gas component at the overall rate of $1 M_\sun$ per year.

\subsection{Amplification and Structure of Magnetic Field}

Magnetic field amplification  originating from the small-scale, randomly oriented dipolar magnetic fields is apparent through the exponential growth, by several orders of magnitude,  of both the magnetic flux and magnetic energy, as displayed in Figure~\ref{fig:mflx-emag}. The growth phase of magnetic field starts at the beginning of the simulation, initially via a quick accumulation of magnetic energy of the supplied magnetic dipoles. The growth of magnetic field strength saturates at about $t = \unit{4}{\Gyr}$, reaching values  $\unit{3-5}{\muG}$ in the disk. During the amplification phase, magnetic flux and total magnetic energy grow by about 6 and 10 orders of magnitude, respectively. As is remarkable in Figure~\ref{fig:mflx-emag}, the growth in the magnetic flux magnitude is close to exponential.  The average e-folding time of magnetic flux is approximately equal $ \unit{270}{\Myr}$, corresponding to the rotation at galactocentric radius equal approximately $\unit{10}{\kpc}$. During the initial period ending at $t=\unit{2.3}{\Gyr}$,  magnetic flux changes sign and its absolute value variates randomly around the exponential curve. These variations are associated with the evolving magnetic field structure  shown in Figure~\ref{fig:slices-bb}. Magnetic field initially is entirely random, at $t=\unit{20}{\Myr}$, since it originates from randomly oriented magnetic dipoles. Later on the toroidal magnetic field component forms a spiral structure revealing reversals in the disk plane, as is apparent at $t =\unit{700}{\Myr}$.  Magnetic field structure evolves gradually toward larger and lager scales. The toroidal magnetic field direction becomes almost uniform inside the disk around $t=\unit{2.5}{\Gyr}$, when the  total magnetic flux stops to reverse in Figure~\ref{fig:mflx-emag}. The domination of fluctuating magnetic field, during the transformation phase of magnetic structure, is apparent also through the excess of the total magnetic energy above the exponential curve. The volume occupied by the well-ordered magnetic field expands continuously till the end of the simulation.

Magnetic field  reveals a spiral structure, as  shown in the face-on views, and expansion of the unidirectionally magnetized disk toward larger galactocentric radii is apparent in the edge-on views. The latter effect seems to reflect the overall outflow geometry. Observation of gas and magnetic field fluctuations moving through the computational domain (in animated images  corresponding to vertical slices in Figure~\ref{fig:slices-bb})  reveals a superposition of vertical thickening combined with  radial expansion in outer parts of the disk. 

We note magnetic field reversals apparent at $t=700 \Myr$ and $t=2500 \Myr$, where the central part of the disk toroidal magnetic field has opposite polarity with
respect to the remaining part of the disk. In the snapshot corresponding to $t=4800 \Myr$, the toroidal magnetic field is almost unidirectional in the horizontal cut
through the galactic midplane, but displays a clear reversal in the vertical slice. The apparent magnetic field configuration comprises two regions of opposite
polarity of toroidal magnetic field, therefore, one reversal remains till the end of the simulation. In a more extensive set of simulations we have done till now,
the uniformity of final magnetic field configurations and the number of reversals depends, among other parameters, on the CR diffusion coefficients, which are known
only to an order of magnitude accuracy. If the parallel CR-diffusion coefficient is reduced by an order of magnitude with respect to the value used in the presented
simulation, magnetic field structure, observed over the period of several Gyr, resembles to that presented in Figure~\ref{fig:slices-bb} for $t=700 \Myr$.

To visualize the magnetic field structure in a manner resembling to radio observations of external galaxies, we construct synthetic radio maps of synchrotron radio-emission, using IDL package by \cite{wiatr-08}. We apply standard procedures of integration of stokes parameters {\it I}, {\it Q}, and {\it U} for the polarized synchrotron emissivity, along the line of sight \cite[see e.g.][]{1986tra..book.....R,1994hea2.book.....L}. To compute synchrotron emissivity, we use magnetic field vectors and CR energy distribution. In the computations of synthetic radio maps, we assume that the ratio of CR  electron energy density to CR proton energy density is 0.01, the maximum and minimum energies of cosmic ray protons are $10^{9}$ and $10^{15} \eV$, respectively, and the power index of CR spectrum is $2.3$.   We neglect the effects of synchrotron cooling of CR electrons, and the effects of Faraday rotation.

In Figure~\ref{fig:radiomaps}, we show the polarized intensity of synchrotron emission (color maps) together with polarization vectors. Electric vectors, computed on the basis of integrated Stokes parameters are rotated by $90^{\circ}$, to reproduce the magnetic field direction averaged along the line of sight, under the assumption of vanishing Faraday rotation effects. The polarization vectors, indicating the mean magnetic field direction, reveal a regular spiral structure in the face-on view, and the so-called \textit{X-shaped structure} in the edge-on view. This kind of structures are presented in radio maps of real galaxies \citep{2005mpge.conf..185S,2006A&A...448..133K,2009RMxAC..36...25K}, 
and have been investigated recently in terms of  CR-driven dynamo  by \cite{2009ApJ...693....1O}. A particular similarity can be noticed between our edge-on synthetic radio map and the radio maps of the edge-on galaxies NGC 891  \citep{2009RMxAC..36...25K} and NGC 253 \citep{2009arXiv0908.2985H}.
In the present global model, the X-shaped configuration is an intrinsic property of the magnetic field structure, since it corresponds closely to the X-shaped distribution of magnetic field in the disk and its neighborhood, as shown in Figure~\ref{fig:slices-bb}.

\section{Discussion and conclusions}

We have shown that the Cosmic-Ray driven dynamo experiment conducted for a whole
galactic disk, seeded by small-scale magnetic dipoles and cosmic rays supplied in
supernova remnants, amplifies magnetic fields exponentially,  up to the
equipartition level, and develops large-scale  magnetic fields in the disk and the
surrounding galactic halo. The amplification timescale of the large-scale magnetic
field component is close to the disk rotation period, i.e., $\unit{270}{\Myr}$ for
the present simulation. The initially disordered magnetic field contributions from
randomly oriented magnetic dipoles are efficiently ordered within the first
$\unit{2}{\Gyr}$ of disk evolution.  The presented experiment supports strongly the
idea that galactic dynamos may have been initiated by small-scale magnetic fields
of stellar origin.

We note, finally, that the present model requires further development, since in the
present version the underlying gravitational potential is axisymmetric. Preliminary
results obtained for simulations with spiral arms indicate the possibility that
reversals are related to the spiral structure in gravitational potential. Our
present model indicates the possibility of efficient action of CR-driven dynamo,
but is still too preliminary to be confronted directly with observational data of a
particular galaxy such as Milky Way. Therefore, incorporation of the effects of
spiral arms is needed for a more detailed  confrontation of our model with the
observational results suggesting existence of magnetic field reversals
\citep[see][and references therein]{2007A&A...464..609H} and magnetic arms
\citep[e.g.][]{2007A&A...470..539B}.

\section*{Acknowledgements}

We thank Harald Lesch for helpful discussions.
The computations were performed on the GALERA supercomputer in TASK Academic
Computer Centre in Gda\'nsk.  This work was partially supported by Polish Ministry of Science and Higher Education through the grants 92/N-ASTROSIM/2008/0 and 3033/B/H03/2008/35.

\bibliographystyle{apj}

\end{document}